\documentclass[%
 aip,
 amsmath,amssymb,
 reprint,%
]{revtex4-1}

 \usepackage{fullpage}
 \usepackage{amssymb}
 \usepackage{amsmath}
 \usepackage{graphicx}
 \usepackage{subfigure}
 \usepackage{dsfont}
 \usepackage{mathrsfs}
 \usepackage{enumerate}
 \usepackage{float}
 \usepackage{dcolumn}
\usepackage{bm}%

\begin{document}
\preprint{AIP/123-QED}



\title{Symbolic partition in chaotic maps}

\author{Misha Chai}%
\affiliation{School of Science, Beijing University of Posts and Telecommunications, Beijing, China 100876}
\author{Yueheng Lan}
\email[Corresponding Author: ]{lanyh@bupt.edu.cn}
\affiliation{School of Science, Beijing University of Posts and Telecommunications, Beijing, China 100876}
\affiliation{
  State Key Lab of Information Photonics and Optical Communications,
 Beijing University of Posts and Telecommunications, Beijing, China 100876
 }



\date{\today}
\begin{abstract}
In this work, we only use data on the unstable manifold to locate the partition boundaries by checking folding points at different levels, which practically coincide with homoclinic tangencies (HTs). The method is then applied to the classic two-dimensional H\'{e}non map and a well-known three-dimensional map. Comparison with previous results is made in the H\'{e}non case and Lyapunov exponents are computed through the metric entropy based on the partition, to show the validity of the current scheme.\\
\end{abstract}



\maketitle

\begin{quotation}
Symbolic dynamics is a very effective description of chaotic motion that captures robust topological features but ignores coordinate-dependent metric properties  of a system.  However, it is difficult to produce a good symbolic partition, especially in high dimensions where the stable and unstable manifolds get entangled in a complex manner. In this paper, we propose a new scheme which only focuses on the unstable manifold and thus avoid the computation of the possibly high-dimensional stable one. With this simplification, the scheme may be applied in more general situations to efficiently carry out the symbolic partitions.
\end{quotation}




\section{Introduction}

\par In the mid-1600s, Newton started the research on differential equations and solved the two-body problem. Later-generation mathematicians and physicists tried to extend Newton's method to the well-known three-body problem but miserably failed until Poincar\'{e} introduced a new point of view which focuses on qualitative rather than quantitative features of the dynamics. When a motion turns chaotic, analytic approximation becomes less useful and geometric description seems more natural. A very useful tool to represent the topological feature of chaos is symbolic dynamics, which translates points on the attractor into long sequences of symbols drawn from a given set labeling different patches of the attractor, and the dynamics into a shift in the symbol sequence. The key to the construction of good symbolic dynamics is to find a simplest partition which is able to assign each point on the attractor a unique symbol sequence\cite{collet2009iterated}. \\

\par In flows,  although it is possible to distinguish different orbits with orbit topology \cite{dong2014}, a common practice is converting it to a map on a well-chosen Poincar\'{e} surface of section. If a 1D map is chaotic on an interval, it is possible to make partitions with extremum points \cite{Chaosbook}. In higher dimensions, things become much harder and we need to check the homoclinic tangencies (HTs) of the stable and unstable manifolds of particular invariant set. For some two-dimensional maps, many approaches which heavily rely on the geometry of phase space, have been successfully used to generate symbolic partitions. H\'{e}non map \cite{Henon1976}, for example, a classical two-dimensional map, is used by many authors for different schemes of symbolic partition \cite{giovannini1992generating, 1985Generating, 1993Hansen, 1989Kantz, 1997Kantz, 1999Politi, cvitanovic1988, kantz1997improved, politi1994symbolic}, where the stable and  unstable manifold of a fixed point is often built to search for the HTs, both of which are one dimensional and relatively easy to compute. Even so, at some parameter values, the precise determination of the primary homoclinic tangencies (PHTs) turns illusive \cite{giovannini1992generating, 1985Generating, 1993Hansen, 1989BW}. For maps in three or more dimensions, the partition becomes even harder since the stable or unstable manifold has a dimension higher than one, which may be very difficult to describe quantitatively. In Hamiltonian systems or the like, an interesting homotopic lobe dynamics could be used to define a symbolic partition, where ``holes" play an essential role instead of the HTs in the usual consideration\cite{mitchell2012partitioning, mitchell2009topology, collins2002symbolic, collins2005forcing, easton1986trellises, rom1994homoclinic, ruckerl1994scaling}. One interesting set of approaches rely on unstable periodic orbits densely embedded in the chaotic attractor, to generate unique symbol sequence\cite{1989BW, badii1994progress, Biham1990, Cvitanovic1991, davidchack2000estimating, plumecoq2000template, plumecoq2000template2}. However, in order to guarantee the accuracy, those methods require a sufficient number of unstable periodic orbits which is a challenge in the case with limited, noisy, time series data. Symbolic partition could also be constructed with the help of a network that well shapes the system dynamics and sometimes approximation of the generating partition could be obtained by properly designed stochastic optimization techniques\cite{kennel2003estimating, buhl2005statistically, patil2018empirical, gao2009complex, donner2010recurrence, nakamura2016constructing,  sakellariou2019markov}.\\


\begin{figure}
\centering
\includegraphics[width=8.cm]{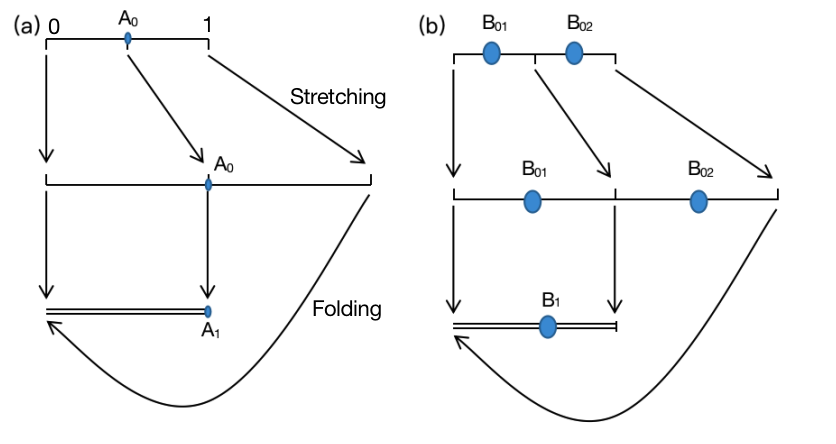} 
\caption{\label{fig1} The mapping of points in the process of stretching and folding: (a) point $A_0$ will be mapped to point $A_1$ after one iteration; (b) the points in area $B_{01}$ and $B_{02}$ will be mapped to area $B_1$ after one iteration. }
\end{figure}

\par Here we propose a new approach which focuses only on unstable manifolds of maps. Since real-world maps including those defined on Poincar\'{e} sections for flows are quite dissipative, {\em i.e.}, the Lyapunov dimension is low. In fact, the unstable manifolds of most well-known maps have a dimension less than three and often are even one-dimensional in some parameter regime, which indicates a high dimension of the stable manifold and thus brings trouble to the conventional computation of HTs. In our method, however, folding points could be conveniently determined on the unstable manifold, which could be identified as a subset of the HTs and thus used for symbolic partition. For finite resolution, the set of partition points once determined can also be used to deduce the number of symbols that are needed. The scheme is tested on the H\'{e}non map at different parameter values and successfully applied to a well-known three-dimensional map.\\

\par In the following, we explain the basic idea of the new approach based on an observation in the 1D map in Section 2, emphasizing the importance of the stretching and folding mechanism of chaos generation. In Section 3, H\'{e}non map is used as an example to show how partition points are detected on the unstable manifold of a fixed point, when foldings are strong or not so strong. To further check the validity of the new method, we will extend the application to a three-dimensional map in Section 4, which has a very complex attractor, being proposed by A. S. Gonchenko and S. V. Gonchenko \cite{2016Gonchenko, 2014Gonchenko} and also mentioned in Ref \cite{2005Gonchenko}. Compared with 2D maps, new troubles emerge concerning the 3D structure, but our method still works very well. In the end, we compute the metric entropy to justify our partition, like what other authors did \cite{1985Generating, 1989Kantz, 1997Kantz, 1987Entropy, 1984Badii}. The results are summarized in Section 5.


\section{Folding points and symbolic partition}

\par First, we have a close look at the folding points that determine  the symbolic partition in the 1D map. As we know, the action of a map leading to chaos consists of two steps \cite{ChaosOtt}: stretching and folding. For example, in the logistic map $x_{n+1} =rx_n(1-x_n)$ with $r=4$, the points in the interval [0, 1] can be viewed as being stretched twice to its original length and then folded back as shown in FIG.~\ref{fig1}. As a result, the interval [0, 1] is decomposed into two subintervals on either of which the map is monotone and extends the subset to the full interval. As the iteration goes on, each subinterval is divided into smaller and smaller intervals to ensure the monotonicity. In general, our method starts with a ``baseline''  which plays a similar role to the interval [0, 1] but needs to be determined (see FIG.~\ref{fig1} and Remark \uppercase\expandafter{\romannumeral2} below), which may not be so obvious in high-dimensional maps, especially when the folding is not strong enough. Iterations of the baseline leads to layers of segments, each of which looks more or less similar to the baseline and is sequentially connected to each other with ``folding points" to be characterized in detail below. Therefore, each segment is bounded by two folding points at the two ends and can be viewed as some kind of ``maximally stretched" piece of the manifold.\\

\par Hence, precisely determining the location of foldings is essential to the symbolic partition since in the process of stretching and folding, two areas $B_{01}$ and $B_{02}$ will be mapped to the same area $B_1$ after one iteration as shown in FIG.~\ref{fig1}, so that the symbolic sequences after one shift would be the same for the corresponding points in $B_{01}$ and $B_{02}$. Therefore, these pairs of points have to lie in different symbol regions before the mapping. Thus the folding point should play the role of the partition point since any neighborhood of it contains points approaching each other after one or several iterations, which belongs to the set of HTs mentioned before in two or higher dimensions. As displayed in FIG.~\ref{fig1}, $A_1$ is the ``folding point" and the partition should be made at its preimage --- the ``critical point'' $A_0$, which is also called ``primary turning point'' in the literature\cite{1993Hansen}. Thus, we have\\

\par {\bfseries Remark \uppercase\expandafter{\romannumeral1:}} folding points emerge from multiple iterations of the baseline and a segment is part of the manifold between two consecutive folding points which is maximally stretched locally. Images of folding points are still folding points and for each genealogy group there is a starting one whose preimage is called a critical point where the radius of curvature is about the size of the attractor.\\

\par To locate a folding point precisely, therefore, we may do a few more iterations in the relevant small neighborhood to get a fully folded structure and unambiguously pick up the unique point with the maximum curvature and then make a few inverse iterations of this point to get it. The key of our method is to define a proper baseline and to locate proper folding points to separate these segments, which is similar to the search for PHTs in other algorithms \cite{giovannini1992generating, 1985Generating, 1993Hansen}. \\
 \par {\bfseries Remark \uppercase\expandafter{\romannumeral2}:} The baseline satisfy the following three conditions: 
\begin{itemize}
\item The chosen fixed point lies on the baseline.
\item The baseline is part  of the unstable manifold of the saddle point.
\item It stretches continuously in both directions until touching the folding points. 
\end{itemize}
In the current scheme, only a well-selected segment on the unstable manifold of a chosen fixed point is employed as the baseline, which will be iterated a few times to get layers of unstable manifold segments that are separated by the folding points. Each folding point defines a family of points and we need to pick up one as the first folding point and the preimages of these folding points can be chosen as the critical points for partition. In the following, if not stated otherwise, the term ``folding point'' is usually referring to the first folding point in its family.\\
\begin{figure*}
\centering
\includegraphics[width=14.cm]{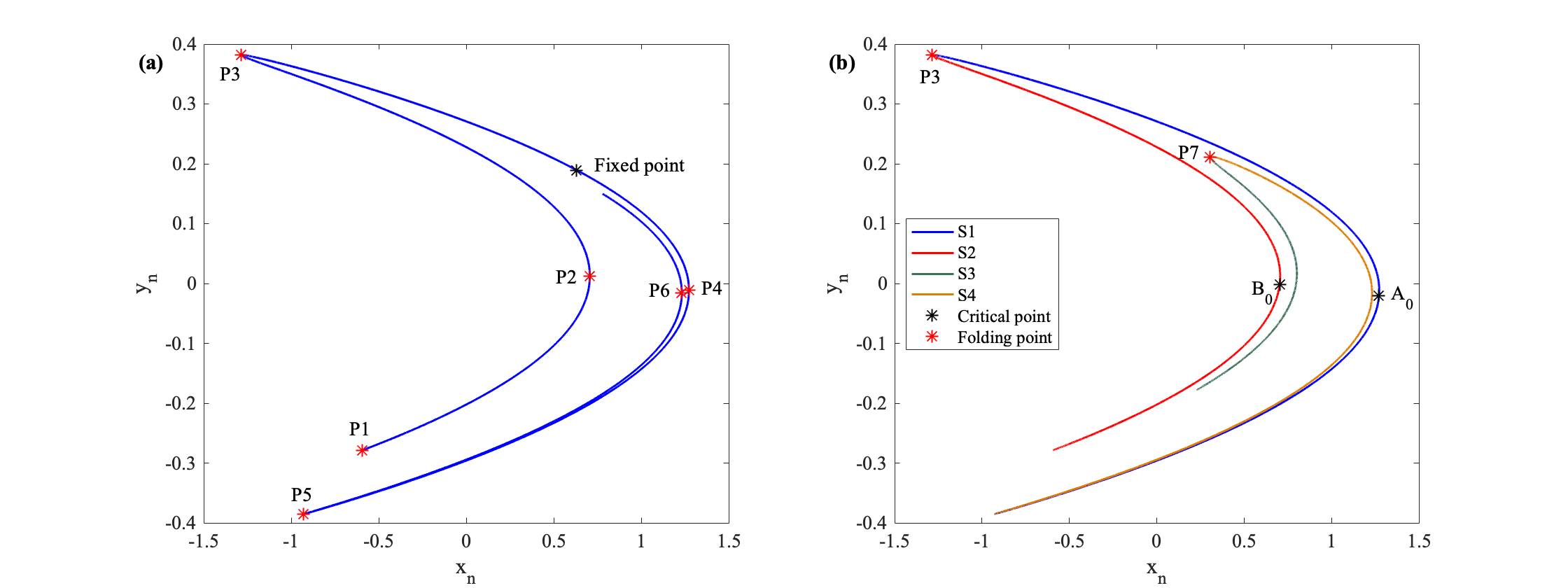}
\caption{\label{fig2} H\'{e}non map Eq.~(\ref{eq:HenonMap}) with a=1.4, b=0.3: (a) part of the unstable manifold of the fixed point (black star) with six maximum curvature points (red star). (b) the folding and the critical points: by one iteration of the baseline S1, a new segment S2 emerges with the folding point P3 and the critical point $A_0$ on S1 is the preimage of P3; one iteration of S2 results in two new segments S3 and S4 along with the folding point P7. The critical point $B_0$ on S2 is the preimage of P7.}
\end{figure*}
 \begin{figure*}
\centering
\includegraphics[width=14.cm]{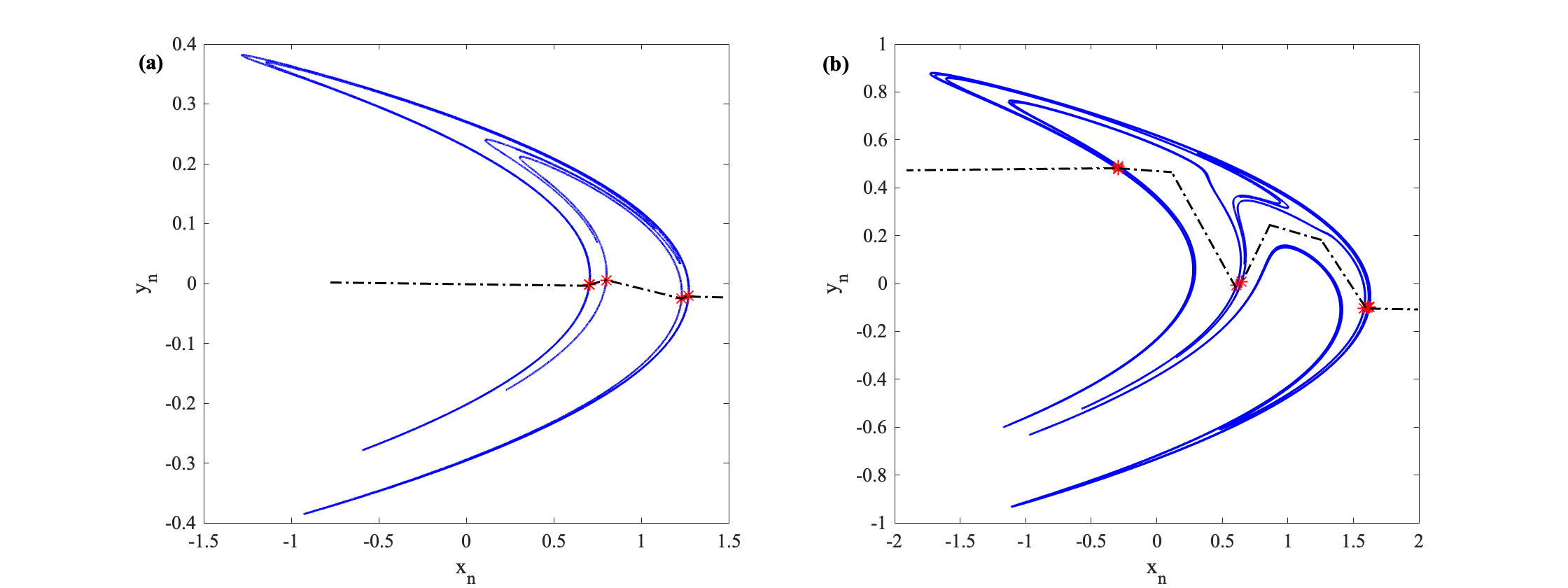} 
\caption{\label{fig3} The symbolic partition for H\'{e}non map at different parameters and the partition line (black dot-dashed line) connects the critical points (red star): (a) $a=1.4, b=0.3$; (b)  $a=1.0, b=0.54$.}
\end{figure*}


\par In addition, we also want to know how many symbols is required for a chaotic map by simply checking the number of critical points on each segment in different layers. For example, in the logistic map with r=4, there is one folding point after one iteration of the baseline, so the baseline can be decomposed to two intervals represented by two symbols, 0 and 1. As later iterations exactly overlap the same interval, there would appear no new critical point and hence two symbols are enough for a good partition. Thus we have\\

\par {\bfseries Remark \uppercase\expandafter{\romannumeral3:}} the number of symbols for a map is determined through the following two steps: \\
\begin{itemize}
\item If the number of critical points is $N$ on a segment, the number of symbols is $N+1$ locally.
\item The number of symbols for a map may be chosen as the maximum number of symbols among all segments.
\end{itemize}

To summarize, for a particular map, our partition scheme consists of the following steps: \\
\begin{enumerate}[(1)]
\item Define a proper baseline in the attractor according to Remark \uppercase\expandafter{\romannumeral2};
\item Iterate the map on the existing part of the unstable manifold and locate the newly emerging folding points according to Remark \uppercase\expandafter{\romannumeral1};
\item Repeat step (2) until a preset resolution is reached;
\item Determine the set of critical points according to Remark \uppercase\expandafter{\romannumeral1};
\item Determine the number of symbols according to Remark \uppercase\expandafter{\romannumeral3}, based on which the symbolic partition is done.
\end{enumerate}

\section{Generating symbolic partition in 2D maps}

\par In this section, we will apply the new scheme to the classical H\'{e}non map for different parameter values. The  equation of the H\'{e}non map is
\begin{eqnarray}\label{eq:HenonMap}
x_{n+1}=-ax_n^2+y_n+1,
\quad
y_{n+1}=bx_n,
\end{eqnarray}
 where $a$ is a parameter that controls the folding and $b$ for the dissipation. With the conventional value $a=1.4, b=0.3$, the folding is strong enough for us to crisply locate the critical points, while with $a=1.0, b=0.54$ where the folding is insufficient, troubles emerge which we will show how to deal with in our scheme. Just like what we did in 1D maps, we first need to find a baseline on the unstable manifold. Hobson \cite{1993Hobson} proposed a numerical scheme which computes stable or unstable manifolds quite accurately and is hence utilized in the following. It starts from a short line near the fixed point along the unstable eigenvector, and an approximation of the unstable manifold results from multiple iterations of this short line.  \\
\subsection{Parameter: a=1.4  b=0.3}

\par In this case, following Hobson's procedure, we choose a short line $ |x-x^*|< 0.001$ along the unstable direction through the fixed point that lies nearest to the attractor, which produces the unstable manifolds in FIG.~\ref{fig2}(a). The structure of the H\'{e}non attractor suggests that the attractor is Cantor-like in the transverse direction and the saddle point sits on the edge of the attractor. Benedicks and Caeleson \cite{1991Benedicks} and Sim\'{o} \cite{1979Simo} proved that the attractor is the closure of the unstable manifold of the saddle point.  \\
 \begin{figure*}
\centering
\includegraphics[width=14.cm]{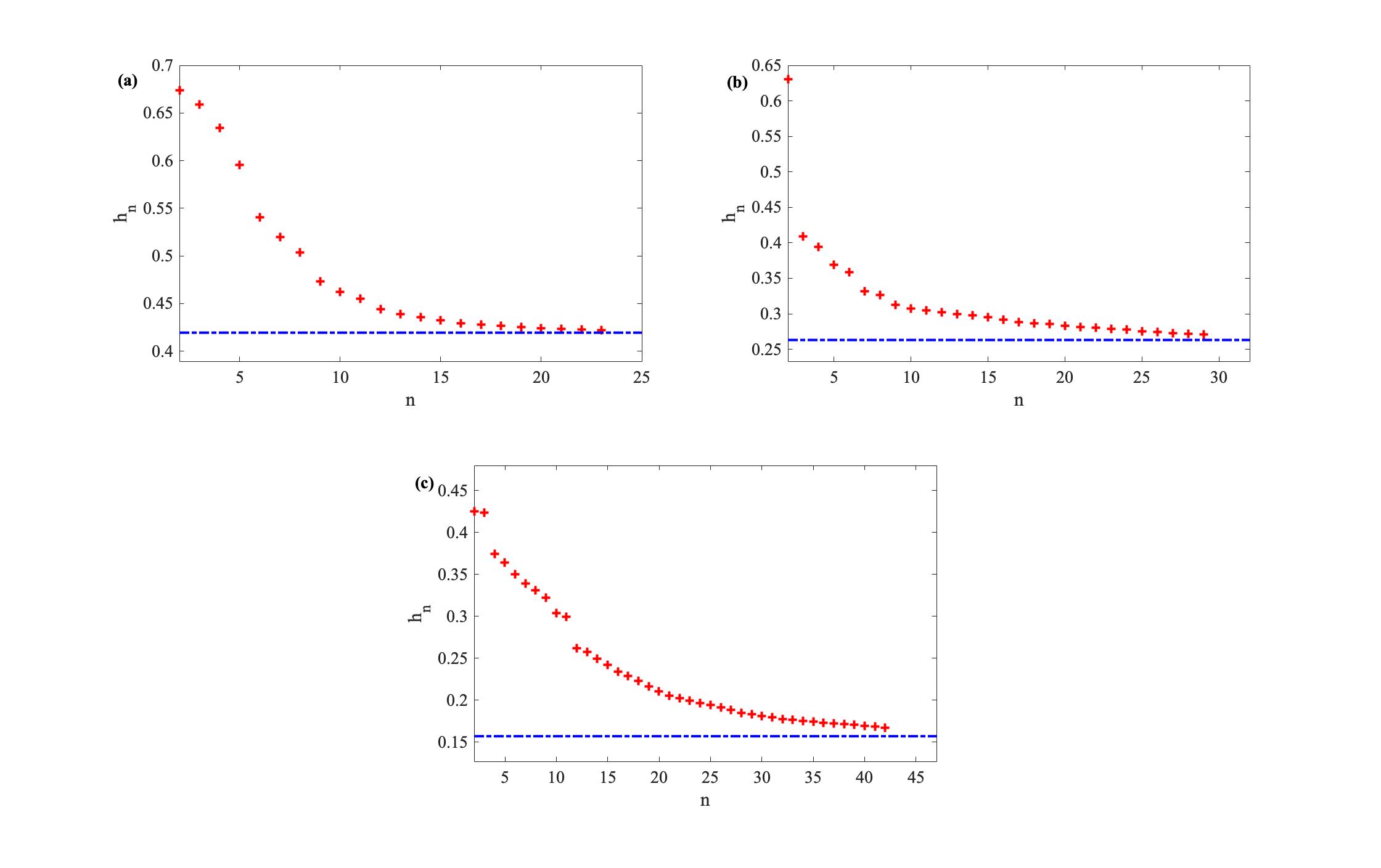}
\caption{\label{fig4}The entropy $h_n$ (red circles) Eq.~(\ref{eq:entropy_h}) for the H\'{e}non map Eq.~(\ref{eq:HenonMap}) with a=1.4 b=0.3 and a=1.0 b=0.54 and 3D map Eq.~(\ref{eq:3DMaps})  with a=-1.86 b=0.72 c=0.03. n denotes the length of the considered symbol sequences. The horizontal blue line indicates the benchmark value of the Lyapunov exponent: (a) for the partition in FIG.~\ref{fig3}(a); (b) for the partition in FIG.~\ref{fig3}(b); (c) for the partition in FIG.~\ref{fig7}.}
\end{figure*}
\par To determine the baseline, we calculate curvatures along the unstable manifold in FIG.~\ref{fig2}(a). The local maximum curvature at P1 is $3.6435\times10^4$; P2: $30.5987$; P3: $1.5630\times10^4$; P4: $31.5920$; P5: $1.5802\times10^5$; P6: $32.1653$. The segment between P3 and P5 fits Remark \uppercase\expandafter{\romannumeral2} because it reaches the maximum stretching in both directions which can thus be defined as the baseline. Nevertheless, P3 and P5 are only approximations of the true folding points. If we want to locate them more accurately, a few more iterations of their neighborhoods will be able to determine more precisely the points with local maximal curvatures. The same number of inverse iterations of these points then gives better location of the foldings. In the following, we carry out this.\\

\par In FIG.~\ref{fig2}(b), by iterating the baseline S1 one step, a new segment S2 is produced together with the folding point P3 according to Remark \uppercase\expandafter{\romannumeral1}. Here the S1 and S2 both reach a maximum stretching locally. To locate P3 more accurately, a point with a maximum curvature of $2.6742\times10^{12}$ is reached by iterating nine more steps and P3 is obtained by nine inverse iterations of this point. The preimage of P3 is the critical point $A_0$ where the radius of curvature is of the same order of the whole attractor. For S2, a similar procedure is followed to locate the folding point P7 and the critical point $B_0$. In this process, two new segments S3 and S4 are emerging, the critical points of which could be detected similarly. Putting all this information together, we get the partition in FIG.~\ref{fig3}(a) at this level.\\
\begin{figure*}
\centering
\includegraphics[width=14.cm]{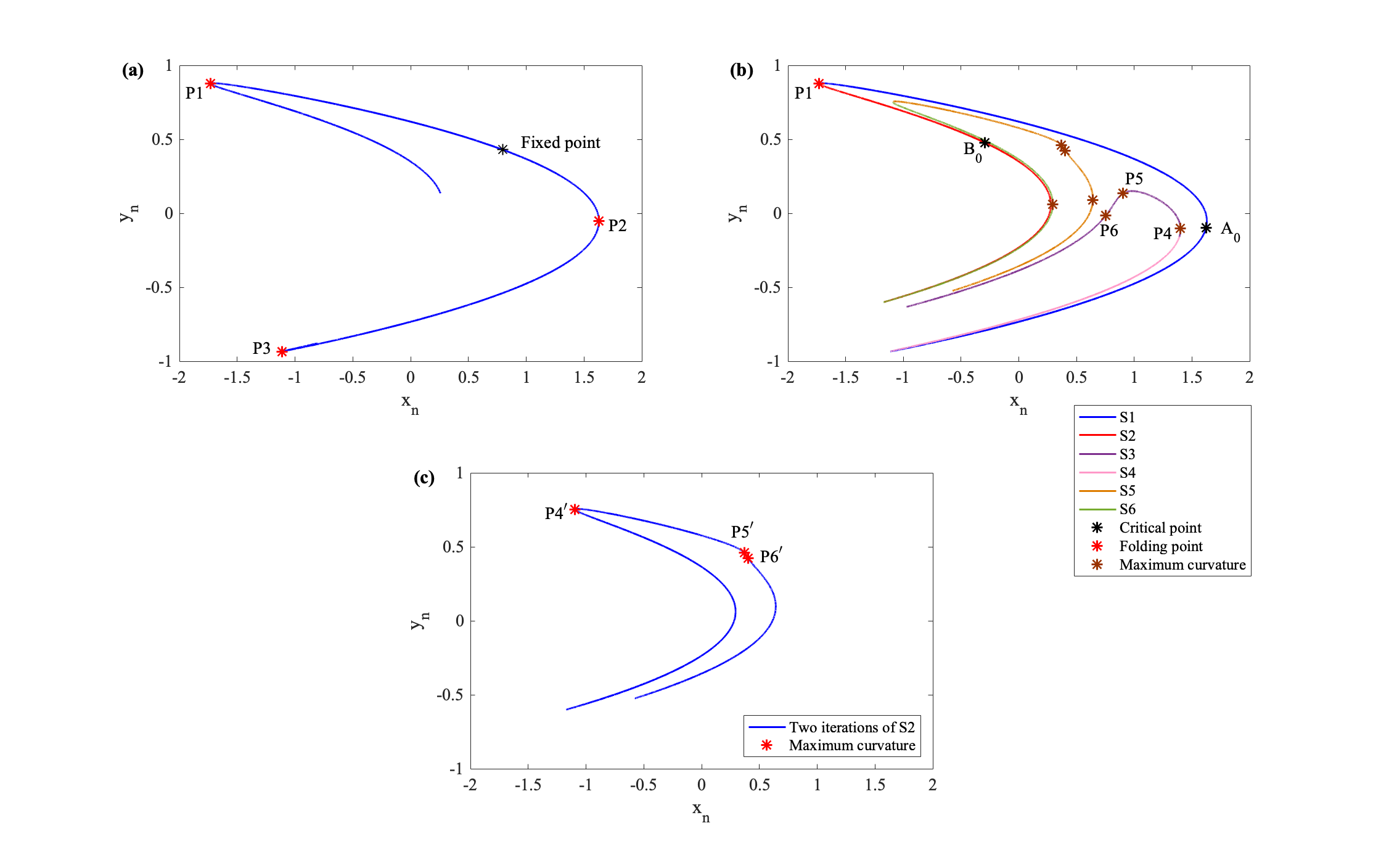} 
\caption{\label{fig5} H\'{e}non map Eq.~(\ref{eq:HenonMap}) with $a=1.0, b=0.54$: (a) part of the unstable manifold of the fixed point (black star) with three maximum curvature points (red star); (b) the folding and the critical points: by one iteration of the baseline S1, a new segment S2 emerges with the folding point P1 and the critical point $A_0$ on S1 is the preimage of P1; one iteration of S2 leads to two new segments S3 and S4 along with the folding point P4. The critical point $B_0$ on S2 is the preimage of P4; one iteration of S3 leads to a new segment S5 along with three maximum curvature points; one iteration of S4 leads to a new segment S6 along with one maximum curvature point; (c) the points P4, P5, P6 will be mapped to points P4$^\prime$, P5$^\prime$, P6$^\prime$ after one iteration.}
\end{figure*}
\par In order to justify our partition, we compute the metric entropy $h$ and compare it with the Lyapunov exponent which is supposed to be equal to $h$ if the partition is correct, as suggested by Grassberger \cite{1985Generating, 1989Kantz, 1987Entropy} and Politi \cite{1984Badii}. Also L. Jaeger and H. Kantz\cite{1997Kantz} showed that for wrong partitions this computation would not have a right convergence. More explicitly, here we use a less biased estimator which was proposed by T. Sch$\ddot{u}$rmann and Grassberger\cite{schurmann1996entropy, schurmann2004bias, schurmann2015note}
 \begin{align}
 \label{eq:entropy_impro}
H_n=\sum_{i=1}^{M}\frac{k_i}{N}(\Psi(N)-\Psi(k_i)+\ln(2)+\sum_{j=1}^{k_i-1}\frac{(-1)^j}{j}),
 \end{align}
where M represents the number of types of sequences $(s_1,...,s_n)$ and $k(s_1,...,s_n)$ the number for each type. $\Psi(x)$  is the logarithmic derivative of the gamma function. 
The difference 
 \begin{align}\label{eq:entropy_h}
 h_n= H_{n+1}- H_n
  \end{align}
should converge to the Lyapunov exponent when $n \to \infty$ if the partition is a valid symbolic one. We choose a trajectory with length $2^{31}$ for the calculation. In this example, with the current setting, the $h_n$ converges well to the Lyapunov exponent $\lambda =0.41938$, as shown in  FIG.~\ref{fig4}(a).\\

\par Here, we would like to mention that the proper folding points are easy to select in the current example since the dissipation of the map is strong enough. In the general case where the dissipation is not enough, ambiguity could arise as in the following example. \\


\subsection{Parameter: a=1.0  b=0.54}

\par Compared to the previous case, here we deal with a situation in which the folding is not strong, leading to uncertainty that entails different partitions \cite{1985Generating, 1993Hansen, 1989BW}. With the current parameter values, Grassberger and Kantz \cite{1985Generating} produced a partition by searching for PHTs, but there is no precise definition on what is ``primary''. Hansen \cite{1993Hansen} arrived at a different partition by employing the same method but also investigating changes of critical points with the parameters. Moreover, he explained why his partition is better. Later various definitions of PHTs were proposed, but may only be used on specific occasions \cite{1985Generating, Grassberger1990}. Giovannini and Politi \cite{giovannini1992generating} used a new method which is also focused on the changes of critical points with the parameters to explore what's characteristic of PHTs and found interesting bifurcations in the generation of symbolic partitions. But for PHTs, they finally gave a conclusion ``... it is not possible to give a priori a nonambiguous definition of the primary homoclinic tangency" and ``We think that the only meaningful way to determine a PHT is via a trial-and-error procedure''. Biham and Wenzel \cite{1989BW} also obtained a different partition through a set of unstable periodic orbits. Grassberger compared his result in Ref \cite{1985Generating} with that in Ref \cite{1989BW}, and delivered an explanation in Ref \cite{1989Kantz}. \\ 
\par As we did before, a few iteration of the short line in the unstable direction of the fixed point results in the structure in FIG.~\ref{fig5}(a), on the initial part of which we find three points with local maximum curvature. The curvature at  P1 is $8.5531\times10^2$; P2: $7.0742$; P3: $1.0226 \times10^5$. According to Remark \uppercase\expandafter{\romannumeral2}, the points between P1 and P3 can be defined as the baseline. In FIG.~\ref{fig5}(b), by one iteration of the baseline S1, a new segment S2 emerges on which we get the folding point P1 according to Remark \uppercase\expandafter{\romannumeral1} and then the critical point $A_0$. Then one iteration of S2 results in two new segments S3 and S4 emerges. However, the folding is not so obvious and it's only mildly folded after one iteration. The curvatures at P4 is 8.2807; P5: 7.3609; P6: 1.3997. Here it's a little bit hard for us to figure out which part will be folded. To determine the exact folding position, we do one more iteration of S2 and see in FIG.~\ref{fig5}(c) that the curve is folded at $P4^\prime$ which thus distills P4 as the folding point. Then the critical point $B_0$ on S2 is the preimage of P4. By one iteration of S3 and S4, the critical points could be detected similarly. However, the difference is, with one iteration of S3, a new segment S5 is produced which is still stretching and no folding point to break it. As a result, no critical point exists on S3. With a similar argument, we conclude that no folding point could be defined on S6 and hence there is no critical point on S4, either. This process could be carried on for more iterations. Finally, after dealing with 180 segments, we get the partition in FIG.~\ref{fig3}(b) and the accuracy reaches $1.0\times10^{-9}$. \\

\par Like what we did before, in order to justify our partition, we compute the metric entropy $h$ which should converge well to the Lyapunov exponent if the partition is a valid symbolic one. The calculation was performed with a series of length $2^{31}$ and the $h_n$ converges well to the Lyapunov exponent $\lambda =0.26315$ with the current setting, as shown in  FIG.~\ref{fig4}(b), which is identical to what K. Hansen obtained in Ref \cite{1993Hansen}.  

\section{Symbolic partition in a 3D map}
\begin{figure*}
\centering
\includegraphics[width=14.cm]{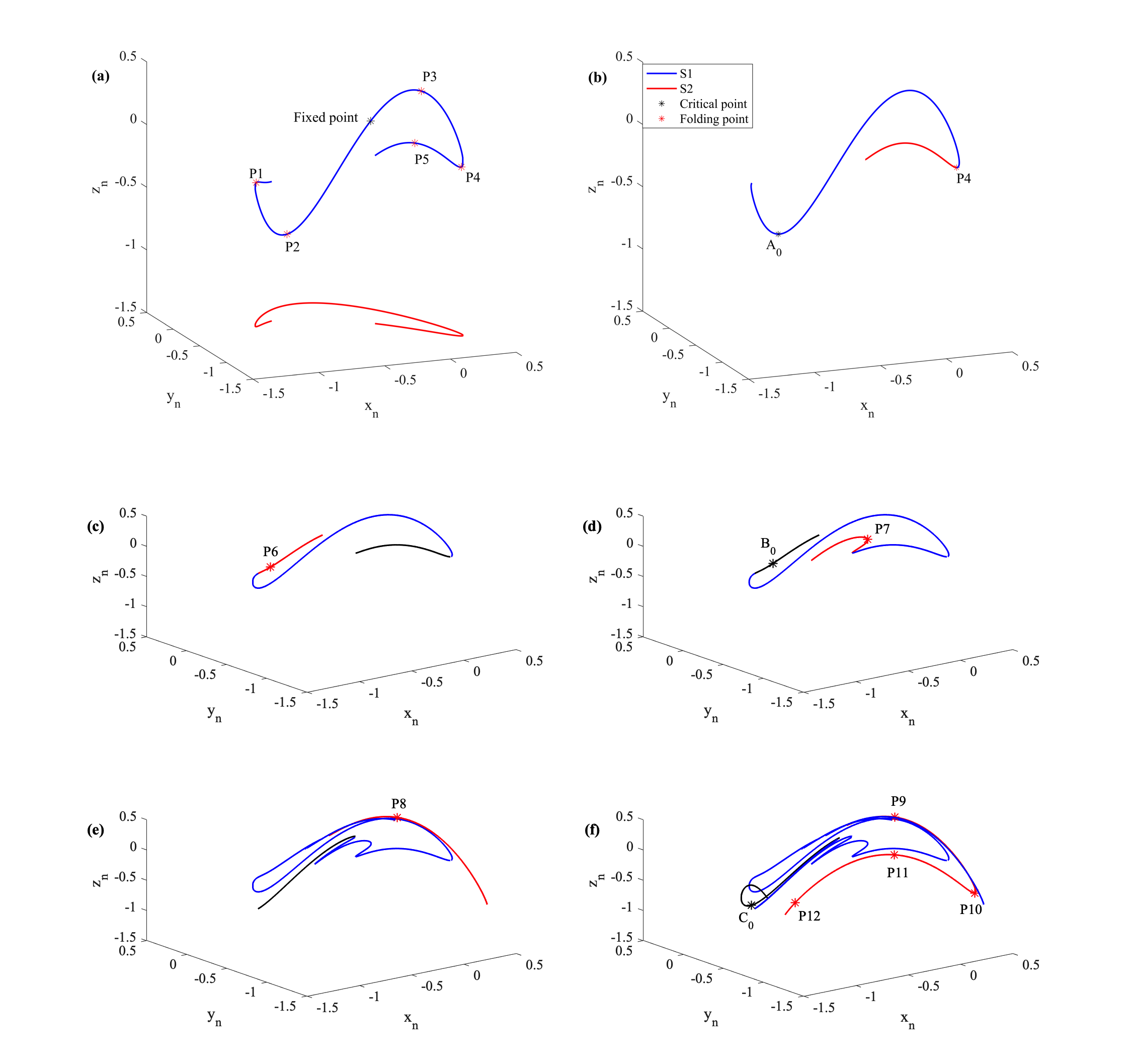}
\caption{\label{fig6} 3D map Eq.~(\ref{eq:3DMaps}) with $a=-1.86, b=0.72, c=0.03$: (a) part of the unstable manifold of the fixed point (black star) with five maximum curvature points (red star). The projection (in red line) on the $(x, y)$-plane is also displayed; (b) the folding and the critical points: by one iteration of the baseline S1, a new segment S2 emerges with the folding point P4 and the critical point $A_0$ on S1 is the preimage of P4; (c) a new red segment emerges with P6 by one iteration of the black segment; (d) two new red segments emerges with folding point P7 by one iteration of the black segment and the critical point $B_0$ is the preimage of P7; (e) a new red segment emerges with P8 by one iteration of the black segment; (f) two new segments emerges with folding point P10 by one iteration of the black segment and the critical point $C_0$ is the preimage of P10.}
\end{figure*}
 \begin{figure*}
\centering
\includegraphics[width=14.cm]{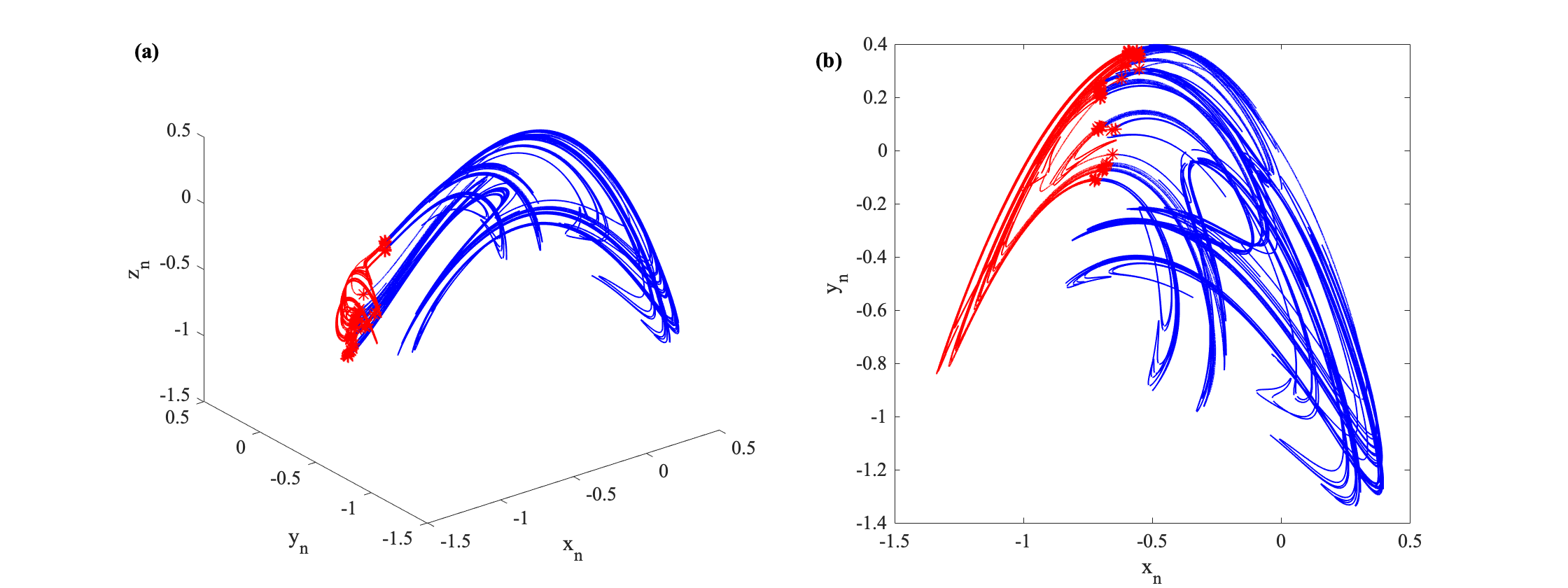}
\caption{\label{fig7} The unstable manifold and the symbolic partition of the 3D map of Eq.~(\ref{eq:3DMaps}): (a) the red curves are given symbol 0; the black curves are given symbol 1; (b) the projection to the $(x, y)$-plane.}
\end{figure*}
\par In this section, we will extend its application to three-dimensional maps. The map that we deal with is proposed by A.S. Gonchenko and S. V. Gonchenkon \cite{2016Gonchenko, 2014Gonchenko}, which could be written as
\begin{align}\label{eq:3DMaps}
&x_{n+1}=y_n  \notag \\
&y_{n+1}=z_n  \notag \\
&z_{n+1}=bx_n+az_n+cy_n-1.45z^2+0.515yz-y^2
\end{align}
where $a=-1.86, b=0.72, c=0.03$. The nonlinear term only exists in the equation for the $z-$component. But the dynamics is chaotic at the current parameter values and the strange attractor looks much more complicated than that of the H\'{e}non map. \\
\par As what we did in 2D maps, we first need to find a proper baseline. However, for 3D maps, it's not as easy as in 2D maps since the unstable manifold appears entangled in a complex way in three dimensions. But Remark \uppercase\expandafter{\romannumeral2} still works. Five points with locally maximum curvatures are identified on the initial part of the manifold as shown in FIG.~\ref{fig6}(a). The curvature at P1 is $178.3072$; P2: $8.3593$; P3: $3.0808$; P4: $76.5475$; P5: $2.7255$. And the iteration goes with $P3\rightarrow P2\rightarrow P4\rightarrow P1$, {\em i.e.}, they belong to the same genealogy group and there is a starting one whose preimage is the critical point. P5 belongs to another genealogy group. According to Remark \uppercase\expandafter{\romannumeral2}, the baseline stretches continuously in both directions until touching the folding points, but at the same time, the baseline between two consecutive folding points is maximally stretched locally according to Remark \uppercase\expandafter{\romannumeral1}. Thus, the baseline captures the overall features of the dynamics but keeps a simple structure. Therefore, the segment between P1 and P4 can be defined as the baseline, which will be folded at P4 after one iteration. According to Remark \uppercase\expandafter{\romannumeral1}, P4 can be defined as a folding point. Its preimage $A_0$ is the critical point shown in FIG.~\ref{fig6}(b). In all the subfigures of FIG.~\ref{fig6}(c)-(f), black segments are mapped to red segments after one iteration. In FIG.~\ref{fig6}(c), there is a maximum curvature of 6.0620 at P6. But because the red segment is still stretching locally and P6 does not break it, so according to Remark \uppercase\expandafter{\romannumeral1}, it can't be defined as a folding point. Therefore, there is no critical point on this black segment. In FIG.~\ref{fig6}(d), the red segment will be folded at P7 with a curvature of 25.0883 and the new segments which are separated by P7 have reached the maximum stretching locally until P7 breaks it, so it can be regarded as a  folding point and thus locates the critical point $B_0$ as a preimage. In FIG.~\ref{fig6}(e), the maximum curvature at P8 is 8.9726 and the red segment is still stretching locally and P8 does not break it, and thus there is no critical point on the black segment for the same reason as in FIG.~\ref{fig6}(c). In FIG.~\ref{fig6}(f), there are four maximum curvatures at P9: 2.9522, P10: 161.9620, P11: 2.9731, P12: 1.0919. P10 can be defined as a folding point  because of the segments which are separated by P10  both reach maximum stretching locally until P10 breaks it. Hence its preimage $C_0$ is the critical point. After dealing with 250 segments, we get the symbolic partition of the attractor as displayed in FIG.~\ref{fig7}.\\
\par Like what we did in 2D maps, in order to justify our partition, we compute $h_n$ and compare it with the Lyapunov exponent if the partition is correct. Here the calculation was performed with a series of length $2^{31}$ and we find that $h_n$ converge well to the Lyapunov exponent $\lambda =0.157$ with the increase of n, as shown in FIG.~\ref{fig4}(c). \\

\section{Conclusion}

\par Symbolic partition is essential for a topological description of orbits in nonlinear systems but remains a challenge for long. In this paper, we focus on the unstable manifold of certain invariant set and carry out the partition based on the stretching and folding mechanism of chaos generation. Three remarks are listed as our guidelines, which starts with the determination of folding points, since folding points not only define the baseline of the manifold but also gives critical points as their preimages. Critical points serve as boundary points for a symbolic partition. Our scheme is successfully demonstrated on the H\'{e}non map with different sets of parameters and on a well-known 3D  map. \\

\par The focus on the unstable manifold in our approach avoids the study of possibly high-dimensional stable manifold, which may accelerate computation in an essential way. As a result, we do not have to search HTs in the full phase space but instead pin down the critical points by iterations only on the unstable manifold. After the iteration genealogy of folding points is sorted out, the partition seems easy to do. However, the determination of the precise starting point in the genealogy could be a problem if the folding process is slow, just as defining the PHTs in the literature which could be a source of confusion~\cite{giovannini1992generating, 1985Generating, 1993Hansen, 1989BW}. Nevertheless, the organization of the layered structure in the current approach may help alleviating difficulties as shown in the examples. \\

\par In the current computation, the determination of the baseline and individual segments is essential to the success of the application. In all the examples, we utilized the unstable manifold of a well-chosen fixed point. Whether this is generally applicable is a question that needs further exploration. Also, we only applied the scheme to maps with just one unstable direction. How to extend it to high-dimensional maps with multiple unstable directions is key for its application in real-world problems. For flows in the phase space, a common practice is to choose a proper Poincar{\' e} section and construct the return map so that the current technique may still apply.  However, in general, it is near impossible to select a good section that works for all orbits and thus a global map is hard to obtain. It appears very rewarding to investigate the possibility of carrying out symbolic partition directly in the full phase space of a flow with an extended scheme.

\section*{Acknowledgement}
This work was supported by the National Natural Science Foundation of China under Grant No. 11775035, and also by the Fundamental Research Funds for the Central Universities with Contract No. 2019XD-A10.   
\section{Data Availability Statement}
The data used to support the findings of this study are available from the corresponding author upon request.



\nocite{*}

\bibliography{manuscript}





\end{document}